\documentclass[aps,nofootinbib]{revtex4}
\usepackage[greek,english]{babel}
\usepackage[applemac]{inputenc}
\usepackage{amstext,amsmath,amssymb,amsfonts,bbm}
\usepackage{enumerate,color}
\usepackage{subfigure}
\def\be{\begin{equation}}
\def\ee{\begin{equation}}
\def\ee{\end{equation}}
\def\ba{\begin{eqnarray}}
\def\ea{\end{eqnarray}}

\def\beta{\rho}

\def\beginitemize{\begin{itemize}\addtolength\itemsep{-8pt}}

\begin{document}

\title{Aristotle's Physics: a Physicist's Look}
\author{Carlo Rovelli}
\affiliation{Aix Marseille Universit\'e, CNRS, CPT, UMR 7332, 13288 Marseille, France.\\
Universit\'e de Toulon, CNRS, CPT, UMR 7332, 83957 La Garde, France.\\
\em email: rovelli@cpt.univ-mrs.fr, tel:+33 614 59 3885.}
\date{\today}
\begin{abstract} 
\noindent
I show that Aristotelian physics is a correct and non-intuitive approximation of Newtonian physics in the suitable domain (motion in fluids), in the same technical sense in which Newton theory is an approximation of Einstein's theory. Aristotelian physics lasted long not because it became dogma, but because it is a very good  empirically grounded theory. The observation suggests some general considerations on inter-theoretical relations. 
\end{abstract}
\maketitle
\section{Introduction}
Aristotle's physics \cite{Aristotle,Aristotle1990,Aristotle1990a} does not enjoy good press. It is commonly called ``intuitive'', and at the same time  ``blatantly wrong". For instance, it is commonly said to state that heavier objects fall faster when every high-school kid should know they fall at the same speed. (Do they??)  Science, we also read,  began only by escaping the Aristotelian straightjacket and learning to rely on observation. Aristotelian physics is not even included among the numerous entries of the \emph{Stanford Encyclopaedia of Philosophy} devoted to Aristotle \cite{Bodnar2012}. To be sure, there are also less hostile and more sympathetic accounts of Aristotle's views on nature, change, and motion, and the historical importance of these views is recognized.   But here is a common example of evaluation:  ``Traditionally scholars have found the notion congenial that Aristotle's intended method in his works on natural science is empirical, even as they have criticized him for failures on this count. The current generation has reversed this verdict entirely. The Physics in particular is now standardly taken as a paradigm of Aristotle's use of dialectical method, understood as a largely conceptual or a priori technique of inquiry appropriate for philosophy, as opposed to the more empirical inquiries which we, these days, now typically regard as scientific"\cite{Bolton1995}.  In other words, Aristotle's science is either not science at all, or, to the extent it is science, is a failure.  

From the perspective of a modern physicists such as myself, this  widespread and simplistic dismissal of Aristotelian physics is profoundly misleading.  Taking this (anachronistic) perspective, I argue here that contrary to common claims Aristotle's physics is counterintuitive, based on observation, and correct (in its domain of validity) in the same sense in which Newtonian physics is correct (in its domain).  

Newtonian physics provides an effective conceptual scheme for understanding physical phenomena.  But strictly speaking it is wrong.  For instance, the planet Mercury follows an orbit which  is \emph{not} that predicted by Newtonian physics.  Einstein's theory provides a better description of gravitational phenomena, which predicts  the observed motion of Mercury correctly.  Newtonian theory matches Einstein's theory in a domain of phenomena that includes most of our experience, but our observational precision on Mercury is sufficient to reveal the discrepancy.  This limitation does not compromise the value --practical, conceptual and historical-- of Newton's theory, which remains the rock on which Einstein built, and a theory of the world around us which is still largely used.  

The relation between Einstein's and Newton's theories is detailed in all relativity manuals: if we restrict Einstein's theory to a certain domain of phenomena (small relative velocities, weak gravitational field...), we obtain the Newtonian theory in the appropriate approximation.  Understanding this relation is not an empty academic exercise: it is an important piece of theoretical physics in the cultural baggage of any good scientist.  It clarifies the meaning of relations between different successful theories and sheds light on the very nature of physical theories: we already know, indeed, that Einstein's theory, in turn, has limited domain of validity (it is invalid beyond the Planck scale).   

I show in this note that the technical relation between Aristotle's physics and Newton's physics is of the same nature as the relation between Newton's physics and Einstein's.  To this aim, I reformulate and derive Aristotle's physics in modern terms (to compare Newton and Einstein theory we must start from the second, of course).  Therefore this is not a paper in history of science: I do not look at Aristotle from his own time's perspective, but rather from the perspective of a later time.  Also, I am not interested here in the complex \emph{historical}  developments that lead from ancient to modern physics\footnote{The literature on this is of course immense; I only mention pointers on ancient \cite{Barbour1989}, middle age \cite{Grant2008} and Galilean \cite{Drake2003,Heilbron2012} science.}. Here I compare the two theories of physics that have had the largest and the longest success in the history of humanity, as a contemporary scientist would describe them: in modern technical terms. (We recover Newton's approximation from Einstein's theory  using Einstein's language, not the other way around, of course.)  I think the comparison sheds light on the way theories are related. In the last section I add some general considerations on the nature of scientific progress.  

\section{Brief review of Aristotle's physics}

History of science may have two distinct objectives. The first is to reconstruct the historical complexity of an author or a period.  The second is to understand how we got to know what we know.  There is tension between these two aims. Facts or ideas of scarce relevance for one may have major relevance for the other.  Take the characteristic case of a scientist who has worked a large part of his life on a theory $A$, soon forgotten and without historical consequences, and for a short period on a theory $B$, which has opened the way to major later developments.   The historian working from the first perspective is mostly interested in $A$ and scarcely in $B$.   The historian working from the second perspective is mostly interested in $B$ and scarcely in $A$, because what matters to him is the way future has developed thanks to $B$.  As a scientist of today, I respect the historians working within the first perspective (without which there would be no history at all), but I regret a trend that undervalues the second.  If we want to understand the past we must do so on its own terms, and disregard the future of that past, but if we want to understand the present we better not disregard the past steps that were essential for getting to the present. This is of importance especially for those of us engaged in trying to push ahead the scientific path of discovery today.  We are not much interested in what scientists did wrong, there is too much of that. We are interested in what they did right, because we are trying to copy them in this, not in that.  

From this perspective, I take the liberty to summarize Aristotle's physics using a modern terminology whenever possible.  Aristotle details his physics mostly in three books: ``Physics'' (below referred to as [Ph]) ``On the Heavens'' (below referred to as [He]) and ``On Generation and Corruption''. The first is the book that has given the name to the discipline; it is a profound masterpiece, it discusses Eleatism, the notion of change, the nature of motion, the infinite, space, time, infinite divisibility  \cite{Wieland1970,Ugaglia2012,Bolton1995}. Some of the issues discussed, such as the nature of Time, are still of central relevance today, for instance in quantum gravity research. But it is not on this I focus here.  The second is simpler and contains most of what we call Aristotle's physics today. I focus here on the parts of the theory that are comparable to Newtonian physics, and which form the basis of the Aristotelian theory of local movement (\foreignlanguage{greek}{for\'a}).  The theory is as follows. There are two kind of motions 
\beginitemize
\item[(a)] Violent motion, or unnatural [Ph 254b10], 
\item[(b)] Natural motion  [He 300a20]. 
\end{itemize}
Violent motion is multiform and is caused by some accidental external agent. For instance a stone is moving towards the sky because I have thrown it.  My throwing is the cause of the violent motion.   Natural motion is the motion of objects left to themselves. Violent motion is of finite duration. That is: 
\beginitemize
\item[(c)] Once the effect of the agent causing a violent motion is exhausted, the violent motion ceases.  
\end{itemize}
To describe natural motion, on the other hand, we need a bit of cosmology. The cosmos is composed by mixtures of five elementary substances to which we can give the names Earth, Water, Air, Fire [He 312a30], and Ether. The ground on which we walk (the ``Earth") has approximate spherical shape. It is surrounded by a spherical shell, called the ``natural place of Water'', then a spherical shell called ``natural place of Air'', then the ``natural place of the Fire'' [He 287a30].  All this is immersed in a further spherical shell  [He 286b10] called the Heaven, where the celestial bodies like Sun, Moon and stars move.  The entire sphere is much larger than the size of the Earth, which is of the order of 400 thousand stadii [He 298a15] (a bit too much, but a correct order of magnitude estimate).  The entire cosmos is finite and the outmost spherical shell rotates rapidly around the central Earth.   Given this structure of the cosmos, we can now describe natural motion.  This is of two different kinds, according to whether it is motion of the Ether, or motion of one of the four elements Earth, Water, Air and Fire. 
\beginitemize
\item[(d)]  The natural motion of the Ether in the Heavens is circular around the center [He 26915].   
\item[(e)]  The natural motion of Earth, Water, Air and Fire is vertical, directed towards the natural place of the substance [He 300b25]. 
\end{itemize}
Since elements move naturally to their natural place, they are also found mostly at their natural place.\footnote{Which, by the way is the source of Aristotle criticism to Anaximander's or Plato's explanation of why the Earth does not move. It is not because ``of indifference" as (at least according to Aristotle), these author claim. Rather, it is because the } 

This is the general scheme. More in detail, Aristotle discusses also the rate at which natural motion happens. He states that
\beginitemize
\item[(f)] Heavier objects fall faster: their natural motion downwards happens faster [Ph 215a25, He 311a19-21]; 
\item[(g)] the same object falls faster in a less dense medium [Ph 215a25]. 
\end{itemize}
Quantitative precision is not very common in Aristotle, who is interested in the causal and qualitative aspects of phenomena.   But in the text following [Ph 215a25], Aristotle uses a mathematical (geometrical) notation from which one can infer that he is actually saying with a certain technical precision that the speed $v$ of fall is proportional to the weight $W$ of the body and inversely proportional to the density $\rho$ of the medium. In modern notation,
\be
{\rm (h')}   \hspace{6.5cm} v\ \sim \ c\ \frac{W}{\rho}.  \hspace{8cm}
\label{vacuum}
\ee
where $c$ is a constant. What one can deduce from Aristotle's discussion is indeed a bit weaker: essentially that the speed would go to infinity if the density of the fluid would go to zero. In modern (and now definitely very anachronistic) terms this could be formulated as
\be
{\rm (h)}   \hspace{6.5cm} v\ \sim \ c \left(\frac{W}{\rho}\right)^n.  \hspace{7.2cm}
\label{vacuum2}
\ee
with positive $n$. About the constant $c$, Aristotle says that 
\beginitemize
\item[(i)] The shape of the body [...] accounts for their moving faster or slower [He 313a14];
\end{itemize}
that is,  the constant $c$ is depends on the shape of the body.\footnote{I am perhaps a bit understating here the variety of Aristotle's attempts to supply principles of proportion for motion, and for speed.}  The context in which Aristotle refers to these relations is a discussion on the void.  Aristotle argues that \eqref{vacuum}, or \eqref{vacuum2}, imply that 
\beginitemize
\item[(j)] In a vacuum with vanishing density a heavy body would fall with infinite velocity [Ph 216a]. 
\end{itemize}
In fact, it is mostly on the basis of this deduction that one can reconstruct \eqref{vacuum2}. On the basis of this (and other) arguments, Aristotle concludes denying the possibility of void:
\beginitemize
\item[(k)] ``From what has been said it is evident that void does not exist [...]" [Ph 217b20]. 
\end{itemize}
In an early dialog \cite{Galilei1960}, Galileo, disliking this conclusion, suggests that it can be avoided by replacing the inverse dependence of $v$ on $\rho$ with a difference (see \cite{Heilbron2010} pg 51), something like $v\sim cW-\rho$, which would avoid the infinite speed in vacuo where $\rho$ vanishes.\footnote{Galileo praises himself for this stupid idea: ``Oh! Subtle invention, most beautiful thought! Let all philosophers be silent who think they can philosophize without a knowledge of divine mathematics!'' Later in life he will make better use of the mathematics that Aristotle lacked.}
 
Two comments before proceeding.  First, Aristotle's choice of \emph{four} elementary substances is strictly dependent on his theory of motion and is deduced from observation. If all things fell down, only one substance would be needed; but some things, like fire, move up. If there were only things moving upwards (like fire) or downward (like earth), two elementary substances would suffice: one with a natural tendency moving upward and one with a natural tendency moving downward. But observation teaches us that there are objects that move upwards in a medium but downward in another. Air bubbles up in water, but is pushed down by up going fire. Wood moves down in air and up in water.  This requires a complex theory or \emph{relations} between several elements [He 269b20-31 and 311a16-b26]. 

Second, contrary to what sometimes stated, the distinction between natural and violent motion survives in later theories of motion.  For instance, the first two laws of Newton clearly reproduce this distinction: in Newton theory, the \emph{natural} motion of a body is rectilinear and uniform (constant speed and straight): this is how a body moves if nothing acts on it.   While \emph{violent} motion is the accelerated motion of an object subject to a force.  The two theories differ in the identification of the ``natural'' motion (rectilinear uniform in Newton, vertical and ending at the natural place in Aristotle), but also in the effect caused by an agent: an external agent causes an acceleration in Newton's theory, while it causes a displacement in Aristotle's theory. But the  fundamental inertial/forced distinction is taken from Aristotle's natural/violent distinction (more on this later).

\section{The Approximation}

Aristotle's physics is the correct approximation of Newtonian physics in a particular domain, which happens to be the domain where we, humanity, conduct our business. This domain is formed by \emph{objects in a spherically symmetric gravitational field (that of the Earth) immersed in a fluid (air or water)} and the main celestial bodies visible from Earth.  The fact that Aristotelian physics (unlike that of most of his commentators) is to be properly understood as the physics of objects immersed in a fluid, air or water, has been emphasized by Monica Ugaglia \cite{Ugaglia2004,Ugaglia2013} and in my opinion is the key to understand Aristotle's physics in modern terms.   

For a student who has learned physics in a modern school it may sound strange to start physics by studying objects in a fluid.  But for somebody who hasn't it may sound strange not to: everything around us is immersed in a fluid.   Aristotle's physics is a highly nontrivial correct description of \emph{these} phenomena, without mistakes, and consistent with Newtonian physics, in the same manner in which Newtonian physics is consistent with Einstein physics in its domain of validity (see also \cite{Moody1975}).  

To see this, we must start by distinguishing the Heavens and the Earth.  Let us start from the Earth.  The domain of terrestrial phenomena in which Aristotle is interested is definitely nonrelativistic and nonquantistic, and therefore we can disregard relativity and quantum theory and start from Newton theory.  Second, Aristotle is interested in movements of objects on the surface of the Earth, both in water and outside water, in air.  The motion of an object in this context is described in Newtonian theory by the equation
\be
\vec F=m\vec a
\ee
where $m$ is the mass of the object,  and $\vec a$ is its acceleration. According to Newton theory, the force $\vec F$ acting on the object is composed by various components that can be simply added. These are: gravity, buoyancy, fluid resistance, plus any other additional force. They are given by the following expression, 
\be
\vec F=-G\frac{m M}{r^2} \vec z + V\rho \vec z - C\rho |v|\vec v +\vec F_{ext}.
\label{due}
\ee
The first term is the force of gravity of the Earth: $G$ is Newton constant, $M$ the mass of the Earth,  $r$ the distance from the center of the Earth and the vector $\vec z$ is the unit vector toward the upper vertical. Since the range of variability of $r$ is small with respect to $r$ for the bodies we are concerned with, we can approximate this term by 
\be
-G\frac{m M}{r^2} \vec z \sim - m g \vec z
\label{approx}
\ee
where $g$ is Galileo Acceleration: $g \sim 9.8\ m/s^2$. The second term is the (Archimedes) buoyancy force due to the weight of the fluid in which the body is immersed; it is different in air and in water; $V$ is the volume of the body and $\rho$ is the density of the fluid.  The third term is the dissipative force due to the resistance of the fluid (water or air) in which the body is immersed; $\vec v$ is the velocity of the body, $C$ is a coefficient that depends on the size and shape of the body.  Finally the last term is the sum of all the forces that are due to other external agents.  The absence of this last term is what Aristotle calls ``natural'' motion as in (b), above. Therefore the distinction in (a) and (b) is simply the distinction between the cases where $\vec F_{ext}$ is present or vanishes.  We deal later with violent motion, for the moment let's stay with natural motion, and therefore have this last term vanish.  

Let's consider a motion which has zero initial velocity.  Its equation of motion at initial time is therefore
\be
m\vec a=-(m g-V\rho) \vec z= (V(\rho-\rho_b)) \vec z
\ee
where 
\be
\rho_b=mg/V
\label{V}
\ee
 is the density of the body.  The body will immediately start moving up or down, according to whether its density is higher or lower than the density of the fluid in which it is immersed. Therefore Earth will move down in any case. Water will move down in Air. Air will move up in water.  Objects that have a specific weight intermediate between water and air (like wood), in Aristotelian terms mixtures including Air as well as Water, will move up in Water and down in Air, and so on. This is precisely the content of (e) above. Furthermore, if a body is immersed in a substance of the same kind, as Water in Water, then it can stay at rest: it is at its natural place.  In other words, the theory of natural motion is the correct description of the vertical motion of bodies immersed in spherical layers of increasingly dense fluids as are the bodies in the domain of validity of Aristotelian theory.

Now let us consider the full natural motion of a body. This if governed by the equation 
\be
m \vec a=-gm \vec z + V\rho \vec z - C\rho |v|\vec v. 
\ee
Assuming for simplicity that the body is initially at rest, we have the one dimensional differential equation 
\be
m\frac{dv}{dt}=-(mg-V\rho)-C\rho v^2. 
\ee
The solution of this differential equation is 
\be
v(t)=\sqrt{\frac{mg-V\rho}{C\rho}}\tanh\left[\sqrt{(mg-V\rho)C\rho}\ t  \right].
\ee
For large $t$ the hyperbolic tangent goes to unity. Therefore the characteristic of this solution is that bodies that fall have \emph{two} regimes: first a transient phase which last for time of the order 
\be
t\sim 1/\sqrt{(mg-V\rho)C}, 
\ee
and then a steady fall where the velocity stabilizes to 
\be
v=\sqrt{\frac{mg-V\rho}{C\rho}}
\ee

The existence of these two phases is important for understanding the common confusion about Aristotle's theory of falling.  Let me explain this key point for the readers less at ease with equations.  A piece of metal falling in water reaches very rapidly a \emph{constant} velocity.  Similarly, a stone left at high altitude by a bird reaches rapidly a \emph{constant} velocity. This true fact of nature is commonly disregarded by most critics of Aristotle's. 

The transient phase during which a body reaches the constant falling velocity is generally too short for a careful observation. For a piece of metal falling in water its duration is often below our ability to resolve it.  For an heavy object (like a stone) falling for a few meters, the time taken to fall is comparable with the transient phase time, therefore the stone does \emph{not} have the time to reach the steady phase. But such a phenomenon implies fast velocities which again are hard to resolve with direct observations (unless one is as smart as Galileo to guess, correctly, that an incline would slow the fall without affecting its qualitative features.)

In most cases of interest the buoyancy term $V\rho$ is negligible with respect to the weight $mg$, and the velocity of fall in the steady regime becomes 
\be
v=\sqrt{\frac{1}{C}}\ \sqrt{\frac{mg}{\rho}}=c \left(\frac{W}{\rho}\right)^{\frac12}.
\label{vv}
\ee
where $c$ is a constant that depends on the shape and the dimension of the body, which is not easy to predict with elementary tools. 

This shows that a heavier body falls faster than a lighter body, precisely as Aristotle states in (f) and that equal bodies fall faster in a less dense medium, as Aristotle states in (g).  The last relation must in fact be compared with Aristotle's relation (h). Finally, at equal weight and density, there is also an effect by the size of the body, as Aristotle states in (i).   We see that Aristotle is perfectly correct in evaluating the falling velocity as something that depends directly on the weight $W=mg$ and inversely on the density of the medium, with a coefficient that depends on the shape of the body. What Aristotle does not have is only the square root, namely $n=\frac12$, which would have been hard for him to capture given the primitive mathematical tools he was using. His factual statements are all correct.  

Hard to claim this is not based on good observation.

If the reader thinks all this is ``intuitive" and ``self-evident", he should ask himself if he would have been able \emph{today} to come up with such an accurate and detailed account of the true motion of falling object. 

Let us now consider violent motion, for terrestrial objects. By definition, these have non vanishing $F_{ext}$. Disregarding for simplicity the weight and buoyancy term, the relevant Newtonian equation of motion is then
\be
m\vec a=- C\beta |v|\vec v+\vec F_{ext}.
\ee
If a body which is initially at rest is subject to a force $\vec F_{ext}$ for a certain time, it will accelerate and reach a velocity $v_o$.  Considering (as does Aristotle) the case when the agent stops acting on the body, the Newtonian equation of motion for the body is then 
\be
m\vec a=- C\beta |v|\vec v
\ee
or, for a motion in one dimension, 
\be
\frac{d^2x}{dt^2}= - \frac{C\beta}m \left(\frac{dx}{dt}\right)^2.
\ee
This is easy to integrate, giving 
\be
x(t)=  \frac{m}{C\beta} \ln\left[v_0 \frac{C\beta}{m}t\right]. 
\ee
where $v_0$ is an integration constant. The velocity is 
\be
v(t)=  \left(\frac{m}{C\beta}\right)\ \frac1t. 
\ee
and goes to zero as the time $t$ grows. The slowing logarithmic growth of $x(t)$ has the consequence that the natural motion brings the object downward before much path can be covered.  This has the effect that any violent motion comes effectively to an end in a finite time, as Aristotle states in (c). 

Let me return to natural motion.  What about the initial transient phase?  Contrary to what many high-school books state, also in this phase the velocity is higher for a heavier body. If the body does not have time to reach its steady state velocity, namely if $t\ll \frac{m}{\beta_f}$ we can estimate the velocity by expanding for small times.   This gives
\be
|v| = \left(g-\frac{V\rho}{m}\right)t , 
\ee
which shows that heavier objects fall faster, precisely as Aristotle states in (f).  The effect is stronger if we keep track of the friction term, of course.  But the fact remains true even disregarding friction!  Heavier objects fall faster even in the approximation where we disregard the friction with the air!

The terrestrial physics of Aristotle matches perfectly the Newtonian one in the appropriate regime.  It is definitely not true that objects with different weight fall at the same speed, in any reasonable terrestrial regime. 

Aristotle's detailed theory however, as well as --seems reasonable-- Aristotle's detailed observations leading to it, refer mostly to the \emph{steady} regime of falling where observation is easier. It disregards the initial transient phase.  This phase is either too short (in water) or too rapid (for very heavy objects in air) for any careful observation.   This phase, on the other hand, is relevant for the short fall of heavy objects, which is the regime on which Galileo (fruitfully) concentrated, circumventing the difficulty of observation by the ingenious trick of the incline.  For this regime, it was already pointed out as early as by Philoponus in the VIth century, that the speed of fall is not  proportional to the weight: a ball of lead doesn't reach ground from a specific height in half the time of ball of half its weight. The buoyancy force and the resistance of the medium do not have the time to become effective in these short falls.  (Two heavy balls with the same shape and different weight \emph{do} fall at different speeds from an aeroplane, confirming Aristotle's theory, not Galileo's.)

Let us now come to the Heavenly physics. Here the regime of interest is that of the bodies we see in the sky, which are not immersed in fluid, are at large distances from Earth and whose apparent motion is slow.   Since they are not immersed in a fluid, we can drop the second and third term from (2).  Since they are distant, we cannot use the approximation \eqref{approx}.    Thus  \eqref{due} becomes now
\be
\vec F=-G\frac{m M}{r^2} \vec z. 
\ee
The simplest solution of this equation (and (1)) is of course given by the circular Keplerian orbits and we know that these happen to describe quite well the relative motions of the Earth-Sun system and the Moon-Earth system.  Since the celestial  bodies are distant and move slowly, we must be careful in translating the motions to our own reference system, which is that of the moving Earth. We must take the motion of the Earth into account.  As well known, to the relevant approximation, the visible motions of stars, Sun and Moon is simply given in Newtonian physics by the apparent rotation of the sky due to the Earth's rotation, the combination of the apparent motion of the Sun due to the Earth's rotation and orbital motion, and the Keplerian orbit of the Moon around the Earth.  All these motions are to a very good approximation ---in fact, exactly so within the observational limits of Aristotle's observational tools---  described by circular motions around the center of the Earth, as in (d).  

We can conclude that Aristotle's physics is correct, in its domain of applicability.  This is given by bodies subjected to a gravitational potential and immersed in fluid (terrestrial physics) and celestial bodies whose motion is either Keplerian around the Earth or the apparent motion due to the Earth's rotation and orbital motion.  Correctly Aristotle distinguishes the two regimes where two different set of laws hold, in the respective approximations, namely (d) and (e). 

Before concluding this technical reconstruction, let us deal with the only two statements we have neglected so far: (j) and (k). The statement (j) follows immediately from equation \eqref{vv}. Therefore it is predicted by the model we are using.  This is at first puzzling: bodies reach infinite speed when falling in vacuum.  The apparent puzzle is resolved by recalling that we have used an approximation.  The relevant approximation here is the one in equation \eqref{approx}. The gravitational force is taken to be constant to derive \eqref{vv} but it is not constant in reality.  A body falling in a hypothetical void is not accelerating forever because at some point it hits the mass originating the attraction.  

What is interesting here is that the infinity is generated by the fact that the theory is approximated. It is corrected by the more complete theory.  This is precisely the expected situation in modern physics with the infinities that appear in general relativity (``singularities'') and in quantum field theory (``ultraviolet divergences''), which are expected to be simply signals that we are using the theory outside its domain of validity.  Therefore Aristotle's deduction (j), and the consequent (k) is correct within the approximation, (as it is correct to say that general relativity yields singularities and quantum field theory ultraviolet divergences) but it is not physically correct because it extrapolates outside the domain of validity of his theory.\footnote{On the other hand, the importance of Aristotle’s conclusion should not, in my opinion, be underestimated. In the ancient atomistic physics of Democritus, the atoms were supposed to move freely in the void; this is consistent with Newtonian inertial motion. But in the later version of this idea developed by Epicurus, it is weight that makes them move. Aristotle had previously shown in the context of his theory of motion that if we extrapolate the common motion due to weight to a situation where the medium has no density at all, as the atomist’s vacuum, the speed of the steady state would be infinite, and the same is true in the Newtonian theory. Therefore Epicurus’s modification of Democritus's inertial motion has a problem which (remember I am taking here the anachronistic perspective of a contemporary physicist) could have been pointed out to him by Aristotle. Democritus's version of atomic free motion is stronger than Epicurus' because it is does not suffer from the problem of an infinite speed that Aristotle correctly deduced.}

In summary, Aristotle's physics of motion can be seen, after translation into the language of classical physics, to yield a highly non trivial, but correct empirical approximation to the actual physical behavior of objects in motion in the circumscribed terrestrial domain for which the theory was created.

\section{Strength and weakness of Aristotle's physics}

Obviously, Aristotle's physics is far from being perfect. In this too it is similar to Newtonian or Einstein's physics, which are far from being perfect either (the first wrongly predicts the instability of atoms, while the second predicts implausible singularities, for example).  Among the various limitations of Aristotelian physics, I illustrate here a few, of different nature. 

\begin{enumerate}
\item According to Aristotelian physics a body moves towards its natural place depending on its composition. This is subtly wrong.  Why does wood float? Because its natural place is lower than Air, but higher than Water.  This was taken in antiquity as the theoretical explanation why boats float. It follows that a boat cannot be built with metal.  Metal sinks. If this theory was correct, metal boats would not float.  But they do. Therefore there is something wrong, or incomplete, in Aristotle's theory.   The point was understood of course by Archimedes: what determines whether or not a body floats in water is not its composition but the ratio of its total weight to its (immersed) volume.  More technically, the quantity $V$ in equation \eqref{V} is not the volume of the body but the overall volume of water it displaces.  This was missed by Aristotle [He 313a15].   Archimedes discovery had major technological and economical consequences \cite{Russo2004}. As soon the true reason for floating was understood, the hull of  Hellenistic kingdoms's ships was covered by a protective metal layer.  This decreased dramatically the need for regular cleaning and therefore the need of pulling the ship periodically out of the water. As a consequence, ships tripled in size (in the III century a.e.v), with a strong impact on trade and development. Theoretical physics has technological and economical consequences also in antiquity. 

\item Aristotle appears to struggle with the distinction between weight and specific weight, without offering a clear distinction between the two. On this, see  \cite{Ugaglia2013}.

\item Violent motion is caused by an external agent.  This is fine.  But Aristotle's premises lead him to assume that the direct effect of the agent stops in the moment it stops acting. This forces him to a complicated and unpalatable explanation of why a stone keeps traveling upward for a while after having left my throwing hand. Aristotle's tentative explanation is based on the effect of surrounding fluid and is  unconvincing.  This led to the medieval theories of impetus and was a major factor for the subsequent advance of physics. The internal difficulties of a good theory are the best hint for advancing our understanding.  The same happened for instance with the equally unpalatable Newtonian action at a distance, which was the key for Einstein's advances. 

\item Aristotle does have an idea of things getting faster and slower but lacks the resources properly to characterize continuous acceleration. This was an issue much considered in mediaeval physics \cite{Grant2008}, but it was Galileo's triumph to understand first empirically (with the incline experiments), then conceptually, the central importance of acceleration, opening the way to Newton's major achievement on which modern physics is built: the main law of motion $F=ma$. 

\item Let me now move to more general methodological concerns. The major limitation of Aristotelian physics, from a modern perspective, is its lack of quantitative developments: Aristotle  is concerned only with the quality, direction, causes, duration of  motion, not the quantitative values of its velocity and so on. Aristotle rarely makes use of mathematics in his science.  Quantitative science was probably stronger in Plato's Academy\footnote{Plato himself, of course, attempted a mathematicization (a geometrization) of atomism in the Timaeus. Beautiful and totally flawed --this is how science often works.  Plato's mistake (from our anachronistic perspective) was to fail to see that to be effective  mathematics had to be used to describe evolution in time, not static shapes.} \cite{Fowler}, for instance with Eudoxus' astronomy, and developed widely in Hellenistic times, especially with Hipparchus, whose marvelously effective mathematical science we know from the Almagest.   

\item There is very little explicit reference to experiments in Aristotelian physics. But this should not be confused with lack of observation.  Aristotelian physics is grounded in accurate observations, like his biology.  One example: a generation earlier, Plato claims to find the idea that the Earth could be spherical reasonable, but says that he would not be able to prove it \cite{Plato1993}.  In his writing, Aristotle provides compelling empirical evidence for this fundamental scientific result, on the basis of a remarkable use of observation: during lunar eclipses, we see the shadow of the Earth projected on the surface of the moon.  By careful observation we see that this shadow is circular [He 297b30].  Notice that there are several geometrical shapes that can project a circular shadow, for instance a cylinder or a cone, but lunar eclipses happen at different hours of the night. In these different situations the Earth is oriented differently with respect to the Sun-Moon line. Therefore it must have a shape that remains circular even if the object is rotated around an axis perpendicular to the direction of  light.  A cylinder and a cone do not have this property, because their shape transforms into a rectangle and a triangle, respectively. The only shape that has this property is the sphere. This proves empirically, and very solidly indeed, that the Earth has a shape which is (approximately) spherical.  It can definitely not be said  Aristotle's physics lacks fine observational ground. 

As much as it lacks of active experimental investigation, Aristotelian physics is rich in deduction.  Several of the arguments Aristotle uses sound wrong to modern ears.  But the strength of Aristotelian deductions in natural science should not be underestimated.  Much of Aristotle's physics is based on observations such as the fact that there are bodies that move  upward in one medium and downward in another, and a rich wealth of consequences that can be deduced from these observations.  Humanity had to wait for Bacon and Galileo to learn the power of directly interrogating Nature, but Aristotelian deduction mode remains in science and has played a major role in the physics of giants such as Einstein and Maxwell. 

\end{enumerate}

A word about the claimed ``intuitive'' aspects of Aristotle's physics.  It is  counterintuitive to think that Earth is spherical and things move vertically in different directions in different parts of the world.  In the 4th century the idea of a spherical Earth was still relatively new and Aristotle provides solid empirical evidence for it in his writings.  The physics compatible with this is far from intuitive.  Aristotle  himself points out the differences between his theory and intuition [He 307b25].  In facts, there are many nonintuitive aspects in Aristotelian physics. The distinction between absolute and relative notions of light and heavy; the idea that the large variety of the things of the word could be accounted for in terms of four elementary substances; the idea that upwards or downwards natural motion stops when the body reaches its natural place; the distinction between natural motion and violent motion, a distinction which, even today, I find hard to understand, in spite of the fact that it remains in Newtonian physics. At the time of Aristotle there were competing physical  schemes, such as those of the atomists, Plato's Timaeus, Empedocles, and I am not aware of any ancient writer that states that the physics of Aristotle is more intuitive than those.   Aristotle goes to great length in criticizing these alternative ideas, using highly non-intuitive arguments.  Aristotle's physics is not intuitive at all. It is a complex and tight conceptual scheme.

Aristotelian physics is often presented as the dogma that slowed the development of science. I think that this is very incorrect. The scientists after Aristotle had no hesitation in modifying, violating, or ignoring Aristotle's physics. Archimedes understanding of the rules of floating is hardly compatible with Aristotelian physics. Ancient astronomy had no hesitation in contradicting Aristotle \cite{Barbour1989}: in his Sun theory, Hipparchus accounts for the difference of duration of the seasons (defined as the time span between equinoxes and solstices) assuming that the Sun orbit is \emph{not} centered on the center of the Earth.  More dramatically, Ptolemy ameliorates Hipparchus predictive system by assuming that celestial bodies do \emph{not} move at constant speed on their path, but rather at a variable speed determined by the equant construction.  This is in flagrant contradiction with Aristotelian physics.  Even in discussing Aristarchus heliocentric ideas, Ptolemy (\cite{Ptolemy1990}, I.7) mentions that they would require a deep revision of Aristotle's physics, but does not seem to consider this as the major obstacle against these idea.  In the Middle Ages the physics of Aristotle was discussed and modified repeatedly, but it took Copernicus, Galileo, Kepler and Newton to find a more powerful theory.  It was not a dogmatic view of Aristotle's theory that kept it alive: it was the difficulty to find something better. In a similar way, Newton theory did not remain the fundamental paradigm for three centuries because it was a dogma, but because it was difficult to find something better. The reason Aristotelian physics lasted so long is not because it became dogma: it is because it is a very good theory.

\section{Incommensurability and continuity}

In my own field of research, theoretical physics, a ``vulgata'' of Kuhn's incommensurability thesis has strong hold.  According to this vulgata, advance in science is marked by discontinuity, the greater the discontinuity the stronger the advance, and not much more than the phenomena survives across the discontinuity.  This has fostered a style of research based on the ideology of discarding past knowledge as irrelevant and working by ``guessing'' possible theories.  In my opinion this ideology is one of the reasons for the current  sterility of theoretical physics.   

Science generates discontinuities and constantly critically reevaluates received ideas, but it builds on past knowledge and its cumulative aspects by very far outnumber its discontinuities.   The Earth was discovered approximately spherical and has remained so; it goes around the Sun and not viceversa, and it will continue to do so; matter has atomic structure, and no Kuhnnian revolution will cancel this; living things on Earth have common ancestors and we are not going to unlearn this... and so on at infinitum. 

Past theories are not cancelled by advanced theories.  They are integrated and better understood within a more powerful perspective.  Einstein's theory does not falsify Newton theory: it clarifies it by neatly specifying its domain of validity, and sheds light on puzzling aspects of the theory by unveiling deeper structures that account for them.   Newton's unpalatable action at distance, for instance, is not cancelled in Einstein's theory: it is simply explained as the approximation in which the finite-speed propagation of the gravitational field is disregarded.  

Advanced theories build heavily on past theories, rebuilding continuously on their conceptual structure and rearranging continuously this conceptual structure. Quine uses repeatedly the beautiful Neurath's boat simile to illustrate this idea \cite{Quine1960}\footnote{``We are like sailors who on the open sea must reconstruct their ship but are never able to start afresh from the bottom. Where a beam is taken away a new one must at once be put there, and for this the rest of the ship is used as support. In this way, by using the old beams and driftwood the ship can be shaped entirely anew, but only by gradual reconstruction."}. 

One can still recognize old Aristotle's vessel, after quite many repairs and improvements, in the conceptual structure of modern theoretical physics.  The Newtonian distinction between inertial motion and motion due to a force, or the modern physics distinction between the kinetic and the interaction terms in the action still are direct traces of the Aristotelian distinction between natural and violent motion.\footnote{I do not think this is merely an analogy, dissenting in this from one of the referees of this paper. Thinking that Aristotelian bodies move naturally as a result of an internal nisus which is the Aristotelian analogue to a force constantly acting on the object is of course possible, but in my opinion it is too much of an anachronism: it is in Newtonian terms that we say that a body falls because of the force of gravity. But to be able to say so, Newton had to tell us what happens to a body over which no force act, which is the purpose of his first law. In other words, he had still to say what is natural motion: a motion on which no force acts. Thus, Newton is still using (and making very good use of) Aristotle's original distinction between natural and violent motion. He simply interprets falling as violent motion, and identifies an agent: gravity.}

This view of the growth of scientific knowledge allows us to talk about past theories in modern terms. Not because this makes the \emph{historical} account of the theory more genuine, of course.  That is, not in view of the first perspective on the history of science.  We do not understand better the historical Newton by knowing that his action at distance is accounted for by general relativity.   But we definitely do better understand our present scientific theories and the historical path that has allowed us to find them, by understanding action at a distance as an approximation.  

From this perspective Aristotle's physics deserves a sharp reevaluation. With all its limitations, it is great theoretical physics. Its major limitation is that it is not mathematical.  Aristotle failed to absorb the Pythagorean visionary faith on the power of mathematics, which Plato recognized and transmitted to his school, from which the great ancient mathematical physics of Alexandria, in particular applied to astronomy, developed.  But Aristotle was able to construct a powerful account of physics which is the ground on which later physics has built.  When Galileo realized that the missing ingredients were the notion of acceleration and the use of formulas, opening the way to Newton, Galileo's interlocutor was Aristotle. Not because Aristotle was the stupid dogma against which intelligence should rise.  But because Aristotle was the best of the intelligence of the world that thirty centuries of civilization had so far produced in this field.  

Of course Galileo, master of propaganda and grand master in the use of words, did his best to ridicule Aristotelian ideas, in the effort to win a difficult battle against a giant.  From this, much of the bad press suffered by Aristotle's physics followed. But Galileo himself, from which so much of the present attitude against Aristotle's physics derives, recognizes the value of the theory of his opponent: he repeatedly opines that Aristotle was enough of an empiricist to modify his view in the light of the new experimental evidence. Indeed, it's a central feature of his rhetoric, to emphasize that his beef is with his contemporary 'Aristotelians' not with Aristotle himself. In a late letter \cite{Galilei}, he writes: ``I am impugned as an impugner of the Peripatetic doctrine, whereas I claim, and surely believe, that I observe more religiously the Peripatetics or should I rather say the Aristotelian teachings than do many that put me down as averse to them.'' And in a letter a month later \cite{Galileia}, he emphasizes the fact that Aristotle put experience before reasoning and concludes ``I am sure that if Aristotle would return to Earth he would accept me among his followers on account of my few but conclusive contradictions to him". 

Galileo's books testify extensively of his struggle with Aristotle's physics in something like a hand-to-hand combat \cite{Drake2003}: several of Galileo's books are punctilious dialogs where one of the character is an Aristotelian. Often Aristotle's idea are  harshly criticized, but those ideas are the ground from which the new Galilean physics starts.  Galileo's struggle with Aristotle is similar to that of Copernicus with Ptolemy, Newton with Descartes, Einstein with Newton or Dirac with Hamilton.  It is building over the past successes and opening them up in depth to modify them that the best science has progressed, in my opinion. Aristotle must be considered as part of the history that has brought us to the present. The continuity between Aristotle and Newton passing by Galileo, is, in my opinion, evident.  Einstein is not even conceivable without the previous work of Newton, Newton is inconceivable without the previous work of Galileo and Galileo is inconceivable without the masterful physics of Aristotle. 

A very recent book  aiming at summarizing the philosophers's doctrines concludes the chapter on Aristotle's physics with the words: ``We can say that nothing of Aristotle's vision of the cosmos has remained valid." (\cite{Natali2014}, page 138.) From a modern physicist's perspective, I'd say the opposite is true: ``Virtually everything of Aristotle's theory of motion is still valid". It is valid in the same sense in which Newton's theory is still valid: it is correct in its domain of validity, profoundly innovative, immensely influential and has introduced structures of thinking on which we are still building.

The bad reputation of Aristotle's physics is undeserved, and leads to widespread ignorance: think for a moment, do you really believe that bodies of different weight fall at the same speed? Why don't you just try: take a coin and piece of paper and let them fall. Do they fall at the same speed?   Aristotle never claimed that bodies fall at different speed ``if we take away the air".  He was interested in the speed of real bodies falling in our real world, where air or water is present. It is curious to read everywhere ``Why didn't Aristotle do the actual experiment?''. I would retort: ``Those writing this, why don't \emph{they} do the actual experiment?''. They would find Aristotle right. 

\vspace{.5cm}

\centerline{------}

\vspace{.5cm}

\noindent Thanks to Monica Ugaglia for very useful advices and criticism, to John Norton for good advice, and to the three anonymous referees whose excellent reports enabled me to much improve the paper.  And warms thanks, come sempre, to Leonard Cottrell. 


\end{document}